\documentclass[11pt, a4paper,twoside]{article}
\usepackage{amssymb,amsfonts,amsmath,amsthm,euscript}
\usepackage[utf8]{inputenc}
\usepackage[english]{babel}
\usepackage[bf,small]{caption}
\usepackage{color}
\usepackage{dsfont}
\usepackage{geometry}
\usepackage{graphicx}
\usepackage{mathrsfs}
\usepackage{mathtools}
\usepackage{multicol}
\usepackage{multirow}
\usepackage{placeins}
\usepackage{setspace}
\usepackage{sectsty}
\usepackage{subfigure}
\usepackage{url}
\usepackage{natbib}
\usepackage{enumerate}

\sectionfont{\large}
\allowdisplaybreaks 

\theoremstyle{plain}

\theoremstyle{definition}

\newcommand{\R}{\mathds{R}}

\newcommand{\Z}{\mathds{Z}}

\newcommand{\bs}[1]{\boldsymbol{#1}}

\newcommand{\F}{\mathscr{F}}
\newcommand{\E}{\mathds{E}}

\long\def\sfootnote[#1]#2{\begingroup%
\def\thefootnote{\fnsymbol{footnote}}\footnote[#1]{#2}\endgroup}

\def\bfootnote{\xdef\@thefnmark{}\@footnotetext}

\begin{document}
\pagestyle{myheadings} 
\markboth{PTSR}{Prass, T.S.; Carlos, J.H.,  Taufemback, C.G. and Pumi, G.} 

\thispagestyle{empty}
{\centering
\Large{\bf Positive Time Series Regression Models}\vspace{.5cm}\\
\normalsize{ {\bf Taiane Schaedler Prass$\!\phantom{i}^{\mathrm{a,}}$\sfootnote[1]{Corresponding author. E-mail: taiane.prass@ufrgs.br.}\let\thefootnote\relax\footnote{\hskip-.3cm$\phantom{s}^\mathrm{a}$Instituto de Matemática e Estatística and Programa de P\'os-Gradua\c c\~ao em Estat\'istica - Universidade Federal Rio Grande do Sul.
} Jonas Hendler Carlos${}^\mathrm{a}$ Cleiton Guolo Taufemback${}^\mathrm{a}$ and Guilherme Pumi${}^\mathrm{a}$
 \\
\let\thefootnote\relax\footnote{This Version: \today}}\\
\vskip.3cm
}}

\begin{abstract}
In this paper we discuss dynamic ARMA-type regression models for time series taking values in $(0,\infty)$.
In the proposed model, the conditional mean is modeled by a dynamic structure containing autoregressive and moving average terms, time-varying regressors, unknown parameters and  link functions. We introduce the new class of models and discuss partial maximum likelihood estimation, hypothesis testing inference, diagnostic analysis and forecasting.
\vspace{.2cm}
\end{abstract}

\section{Introduction}

In the last decades, several models for double bounded time series have been proposed in the literature \citep[][among others]{Fokianos1998,Ferrari2004,Rocha2009,Bayers2017,Pumi2019,PumiFC2019}. These models are usually based on the approach nowadays known as generalized autoregressive moving average models (GARMA). The GARMA approach is based on embedding a time dependent structure into the generalized linear models (GLM) framework. The main idea behind the approach dates back to the late 70's, but the name GARMA was solidified in \cite{Benjamin2003}.

Technically, GARMA models can be categorized as an observation-driven model \citep{Cox1981} and, as such, includes two  main components, the random and the systematic components. On one hand, the random component is responsible for the distributional features of the model, usually depending on a measure of interest, such as the (conditional) mean or median. For instance, in \cite{Rocha2009} the response variable follows a beta distribution, parametrized in terms of its mean, while \cite{Bayers2017} consider a Kumaraswamy distribution parameterized in terms of its median. On the other hand, the systematic component prescribes the dependence structure driving the measure of interest, often called the mean response. For instance, in \cite{Rocha2009}  and \cite{Bayers2017}, both models consider an ARMA-like structure for the systematic component. The difference is that in the former this structure models the conditional mean, while in the later, the conditional median.

GARMA modeling presents several advantages over classical time series models, such as the class of ARIMA models \citep{Box2008}. For instance, GARMA models are tailored to handle bounded non-gaussian time series without the necessity of data transformations, or other adaptations to accommodate such features. Although \cite{Benjamin2003} only consider the case in which the underlying distribution a member of the (canonical) exponential family, several distributions outside the exponential family have been considered in the literature, as for instance, in \cite{Bayers2017}, where the authors consider the Kumaraswamy distribution to model double bounded time series. The distribution applied and the particular structure of the model may vary drastically depending on the characteristics of the data, the application and its goals.

The GARMA framework also allows wide variety of distributional features to be embedded into the model's random component, while retaining any desired dependence structure on the conditional mean response. This leads to a much simpler modeling strategy than the usual approach of inserting distributional features through the error term in linear models also allowing for non-gaussianity to be readily accommodated. Furthermore, conditional inference is naturally accommodated within the framework, providing a powerful inferential tool.

Considering the case where the time series assume only positive values, \cite{Benjamin1998} studies a GARMA model based on the Gamma distribution, while, more recently, \cite{Bea2021} introduces a regression model where the response variable is beta prime distributed. In terms of structure, in both cases exogenous covariates are allowed in the mean response, but the later also includes an ARMA-like structure to handle serial dependence. In this work our goal is to unite these two fronts and propose a class of positive time series regression (PTSR) models allowing the dynamical part of the model to include exogenous (possibly time dependent and random) covariates and also an ARMA-like structure to handle serial dependence for the mean response, in the lines of \cite{Benjamin1998, Rocha2009, Bayers2017, Pumi2019}. Moreover, we shall not restrict ourselves to distributions belonging to the exponential family.

The paper is organized as follows. In the next section we introduce the proposed PTSR model. In Section \ref{s:inf} we introduce a partial maximum likelihood approach for parameter inference in PTSR models and derive closed formulas for the related score vector and Fisher information matrix. Section \ref{s:asinf} we develop the asymptotic theory related to the proposed partial maximum likelihood estimator and from the asymptotic results we derive confidence intervals and hypothesis testing for the model's parameters. In Section \ref{s:diag} we discuss diagnostic analysis and forecast for the proposed model.

\section{Proposed Model}

Let $\{Y_t\}_{t\in \Z}$ be a stochastic process such that $P(Y_t\in(0,\infty)) = 1$ and let $\{\bs X_t\}_{t\in \Z}$ be a set of $s$-dimensional exogenous covariates, possibly time dependent and random.  Let $\F_{t}$ denote the $\sigma$-field representing the history of the model known to the researcher up to time $t$, that is, the sigma-field generated by $(\bs X_{t+1}', Y_t, \bs{X}_t', Y_{t-1},\bs X_{t-1}', \cdots)$. Notice that it is assumed that $\bs{X}_{t+1}$ is known at time $t$. This is always true when $\{\bs X_t\}_{t\in \Z}$ is non-random. In the general case, one can assume that $\{\bs X_t\}_{t\in \Z}$ is a shifted version of a set of covariates and the assumption holds.

Denote by $f(\cdot | \F_{t-1})$ the conditional density of $Y_t$ given $\F_{t-1}$. In this work we only consider distributions for which $\mu_t = \E(Y_t | \F_{t-1})$ exists and is finite with probability 1 and that $f(\cdot | \F_{t-1})$ can be parameterized in terms of $\mu_t$ and, possibly, a parameter $\varphi$ that is not time-dependent. To make this relationship clear, throughout the text we shall set $f(\cdot | \mu_t, \varphi) := f(\cdot | \F_{t-1})$ and use the following notation
  \begin{equation}
  Y_t | \F_{t-1}  \sim  f(\cdot |\mu_t, \varphi), \quad \mbox{where} \quad \mu_t = \E(Y_t | \F_{t-1}), \quad \varphi \in \R.\label{eq:dist_cond}
  \end{equation}

 We propose to model $\mu_t$ using a dynamic GLM-like structure of the form
\begin{align}
\eta_t &:= g_1(\mu_t) = \alpha +\bs{X}_t'\bs\beta + \sum_{k=1}^p\phi_k\bigl[g_2(Y_{t-k}) - I_X\bs X_{t-k}'\bs\beta \bigr]+\sum_{j=1}^q\theta_k e_{t-k},\label{e:gmu}\\
e_t & := Y_t-\mu_t,\nonumber
\end{align}
where $g_1:(0,\infty)\rightarrow \R$ is a twice differentiable, one-to-one link function, $g_2:(0,\infty)\rightarrow \R$ is a link function (not necessarily differentiable),  $\alpha\in \R$ is an intercept, $\bs\beta:=(\beta_1,\cdots,\beta_s)^\prime$ is an $s$-dimensional vector of parameter  associated to the covariates, $I_X$ is an indicator function which assumes the value 1 if the regressors must be included in the AR recursion and 0, otherwise,  $\bs\phi:=(\phi_1,\cdots,\phi_p)^\prime$ and $\bs\theta:=(\theta_1,\cdots,\theta_q)^\prime$ are $p$ and $q$-dimensional vectors of parameters, respectively.

 In \eqref{e:gmu}, $g(\mu_t) = g\big(\E(Y_t|\F_{t-1})\big)$ follows a linear model incorporating the covariates and an adjacent ARMA$(p,q)$-like structure responsible for modeling a possible serial correlation in the conditional mean. In the context of financial time series, $Y_t$ can be viewed as the squared returns while $\mu_t$ is the conditional volatility. The proposed model is observation-driven specified by the random component \eqref{eq:dist_cond} and the systematic component \eqref{e:gmu}.  This approach is closely related to other GARMA-like models for non-Gaussian time series, such as the $\beta$ARMA \citep{Rocha2009}, KARMA \citep{Bayers2017}, $\beta$ARFIMA \citep{Pumi2019} and others \citep[see also][]{Kedem2002}.

 The main difference between the model proposed here and those in the literature is that we consider the error term in the nominal level, namely, $e_t:=Y_t-\mu_t$, instead of the predictive level $e_t:=g_1(Y_t)-\E(Y_t|\F_{t-1})$ considered in the aforementioned works. This difference is mainly due to the fact that $\beta$ARMA and KARMA are models for double bounded time series, while the proposed model considers positive responses. However, it entails an advantage that will be important later: the sequence $\{e_t,\F_{t}\}_{t\in\Z}$ forms a martingale difference sequence. The nature of the positive response is also reflected in the autoregressive component in \eqref{e:gmu}, which is measured in the transformed scale $g_2(Y_t)$.  By choosing different $g_2$ one can either match the linear structure scale $\eta_t$ or keep the original scale (by using the identity function).

\section{Partial Likelihood Inference}\label{s:inf}

Parameter estimation can be carried out by partial maximum likelihood approach. Let $\{(Y_t,\bs{X}_t)\}_{t=1}^n$ be a sample from a PTSR model under specification  \eqref{eq:dist_cond} and \eqref{e:gmu}. Denote by $\bs\gamma:=(\alpha,\bs\beta',\bs\phi',\bs\theta', \varphi)'$ the $(p+q+s+2)$-dimensional parameter vector and let $\Omega\subseteq \R^{p+q+s+2}$ be the parameter space. The partial maximum likelihood estimators (PMLE) are obtained upon maximizing the logarithm of the partial likelihood function given by
\begin{equation}
\ell(\bs\gamma) = \sum_{t=1}^n\ell_t(\bs\gamma), \quad \ell_t(\bs\gamma) := \log\bigl(f(Y_t |\mu_t, \varphi)\bigr),\label{eq:llk}
\end{equation}
so that the partial maximum likelihood estimator of $\bs\gamma$ is given by
\[\hat{\bs\gamma}=\underset{\bs\gamma\in\Omega}{\mathrm{argmax}}(\ell(\bs\gamma)).\]
In most cases, $\bs\gamma$ cannot be analytically obtained and we have to rely on numerical optimization of the partial log-likelihood or upon solving the so-called normal equations.

\subsection{Score Vector}
From \eqref{eq:llk} the derivative of the log-likelihood $\ell(\bs\gamma)$ with respect to $\gamma_j$ is given by
\begin{equation*}
 \frac{\partial \ell(\bs\gamma)}{\partial \gamma_j} =   \sum_{t=1}^{n}\biggl[\frac{\partial\ell_t(\bs\gamma)}{\partial \mu_t}\frac{\partial\mu_t}{\partial\eta_t}\frac{\partial\eta_t}{\partial\gamma_i} +  \frac{\partial \ell_t(\bs\gamma)}{\partial \varphi}\frac{\partial \varphi}{\partial\gamma_i}  \biggr]=  \sum_{t=1}^{n}\biggl[\frac{\partial\ell_t(\bs\gamma)}{\partial \mu_t}\frac{1}{g_1'(\mu_t)}\frac{\partial\eta_t}{\partial\gamma_i}+  \frac{\partial \ell_t(\bs\gamma)}{\partial \varphi}\frac{\partial \varphi}{\partial\gamma_i}  \biggr],
\end{equation*}
so that the score vector $U(\bs\gamma)=\big(U_{\bs\rho}(\bs\gamma)',U_{\varphi}(\bs\gamma)\big)'$, with $\bs\rho:=(\alpha,\bs\beta',\bs\phi',\bs\theta')'$, can be written as
\begin{equation}\label{eq:score}
U_{\bs\rho}(\bs\gamma) = D_{\bs\rho}' T\bs h_1  \qquad \mbox{and} \qquad U_{\varphi}(\bs\gamma) = \bs 1'_n \bs{h}_2,
\end{equation}
where $D_{\bs\rho}$ is the matrix for which the $(i,j)$th elements is given by $[D_{\bs\rho}]_{i,j} = \partial \eta_{i}/\partial \rho_j$, $T$ is a diagonal matrix given by
\begin{align*}
T_1=\mathrm{diag}\biggl\{\frac{\partial \mu_1}{\partial \eta_{t}},\cdots,\frac{\partial \mu_n}{\partial \eta_{n}}\biggr\} = \mathrm{diag}\biggl\{\frac{1}{g_1'(\mu_1)},\cdots,\frac{1}{g_1'(\mu_n)}\biggr\} ,
\end{align*}
$\bs 1_n=(1,\cdots,1)\in\R^n$,  $\bs{h}_1$ and $\bs{h}_2$ are the vectors defined by
\[
\bs{h}_1 = \bigg(\frac{\partial \ell_1(\bs\gamma)}{\partial \mu_1}, \cdots, \frac{\partial \ell_n(\bs\gamma)}{\partial \mu_n}\bigg)' \quad \mbox{and} \quad \bs{h}_2 = \bigg(\frac{\partial \ell_1(\bs\gamma)}{\partial \varphi}, \cdots, \frac{\partial \ell_n(\bs\gamma)}{\partial \varphi}\bigg)'.
\]

Notice that both $D_{\bs \rho}$ and $T$ depend only on the structure defined by \eqref{e:gmu} and will always be the same, independently on the choice of the underlying conditional distribution. Moreover, since $e_t=Y_t-\mu_t$, the following relationship holds
\[\frac{\partial e_{t}}{\partial\gamma_i} = - \frac{\partial \mu_{t}}{\partial\gamma_i}=- \frac{\partial g_1^{-1}(\mu_{t})}{\partial\gamma_i}
=-\frac1{g_1'(\mu_{t})}\frac{\partial \eta_{t}}{\partial\gamma_i},
 \]
and implies that
\begin{align*}
\frac{\partial\eta_t}{\partial\alpha}&=1-\sum_{j=1}^q\frac{\theta_j}{g'_1(\mu_{t-j})}\frac{\partial \eta_{t-j}}{\partial\alpha};\\
\frac{\partial\eta_t}{\partial\beta_i}&=X_{ti}-I_X\sum_{i=1}^p \phi_iX_{(t-i)j} -\sum_{j=1}^q\frac{\theta_j}{g'_1(\mu_{t-j})}\frac{\partial \eta_{t-j}}{\partial\beta_i},\quad i\in\{1,\cdots,s\};\\
\frac{\partial\eta_t}{\partial\phi_i}&=g_1(Y_{t-i})-I_X\bs X_{t-i}'\bs\beta-\sum_{j=1}^q\frac{\theta_j}{g'_1(\mu_{t-j})}\frac{\partial \eta_{t-j}}{\partial\phi_i},\quad i\in\{1,\cdots,p\};\\
\frac{\partial\eta_t}{\partial\theta_i}&= e_{t-i}-\sum_{j=1}^q\frac{1}{g'_1(\mu_{t-j})}\frac{\partial \eta_{t-j}}{\partial\theta_i},\quad i\in\{1,\cdots,q\}.
\end{align*}

Now, upon observing that
\begin{equation*}
 \E\biggl(\frac{\partial \ell(\bs\gamma)}{\partial \gamma_j}\biggr)=  \sum_{t=1}^{n}\E\Biggl(\E\biggl(\biggl[\frac{\partial\ell_t(\bs\gamma)}{\partial \mu_t}\frac{1}{g_1'(\mu_t)}\frac{\partial\eta_t}{\partial\gamma_i}+  \frac{\partial \ell_t(\bs\gamma)}{\partial \varphi}\frac{\partial \varphi}{\partial\gamma_i}  \biggr] \bigg| \F_{t-1}\biggr)\Biggr)
\end{equation*}
and using the fact that ${1}/{g_1'(\mu_t)}$, ${\partial\eta_t}/{\partial\gamma_i}$ and ${\partial \varphi}/{\partial\gamma_i}$ are  $\F_{t-1}$-measurable, one concludes that
\begin{equation*}
\E\biggl(\frac{\partial\ell_t(\bs\gamma)}{\partial \mu_t} \bigg| \F_{t-1}\biggr) = \E\biggl(\frac{\partial \ell_t(\bs\gamma)}{\partial \varphi} \bigg| \F_{t-1}\biggr) = 0, \quad \mbox{implying} \quad  \E\biggl(\frac{\partial \ell(\bs\gamma)}{\partial \gamma_j}\biggr) = 0.
\end{equation*}

\subsection{Conditional information matrix}

In this section we derive the Fisher conditional information matrix, which will be useful later on deriving the asymptotic properties of the partial maximum likelihood estimator for the proposed model.

Let $H_t(\bs \gamma)$ be defined by
\[
H_t(\bs \gamma) = -\frac{\partial^2\ell_t(\bs\gamma)}{\partial \bs \gamma \partial \bs \gamma'},
\]
and observe that
\begin{equation*}
H(\bs \gamma) = -\frac{\partial^2\ell(\bs\gamma)}{\partial \bs \gamma \partial \bs \gamma'}  =  -\sum_{t=1}^{n}\frac{\partial^2\ell_t(\bs\gamma)}{\partial \bs \gamma \partial \bs \gamma'} = \sum_{t=1}^nH_t(\bs \gamma).
\end{equation*}
Also, observe that both, $H(\bs \gamma)$ and $\ell(\bs\gamma)$ depend on $n$, however, for simplicity and since no confusion will arise, we omit this dependence from the notation.

Let $I_n(\bs \gamma) := \E(H(\bs \gamma))$ be the information matrix corresponding to the sample of size $n$ and $I^{(n)}(\bs \gamma)$ is the negative expectation of the hessian $H_t(\bs \gamma)$ averaged over all observations, that is,
\[
   I^{(n)}(\bs \gamma) = -\frac{1}{n}\sum_{t = 1}^n \E \bigg(\frac{\partial^2\ell_t(\bs \gamma)}{\partial \bs\gamma \partial \bs \gamma'} \bigg).
\]
Hence,
\[
I^{(n)}(\bs \gamma) = -\frac{1}{n}\E \bigg(\frac{\partial^2\ell(\bs \gamma)}{\partial \bs \gamma \partial \bs \gamma'} \bigg) \quad \mbox{and}  \quad I_n(\bs \gamma ) = nI^{(n)}(\bs \gamma).
\]

Now, observe that
\[
 I^{(n)}(\bs \gamma) = -\frac{1}{n}\sum_{t = 1}^n \E \bigg( \E\bigg(\frac{\partial^2\ell_t(\bs \gamma)}{\partial \bs\gamma \partial \bs \gamma'} \bigg| \F_{t-1}\bigg)\bigg)  = \frac{1}{n}\E(K_n(\bs \gamma))
\]
with
\[
 K_n(\bs \gamma) : = -\sum_{t = 1}^n \E \bigg(\frac{\partial^2\ell_t(\bs \gamma)}{\partial \bs\gamma \partial \bs\gamma'} \Big| \F_{t-1}\bigg).
\]
The matrix $K_n(\bs \gamma)$ is known as the conditional information matrix corresponding to the sample of size $n$ and its $(i,j)$th element is given by
\[
[K_n(\bs \gamma)]_{i,j} = -\sum_{t = 1}^n \E \bigg(\frac{\partial^2\ell_t(\bs \gamma)}{\partial \gamma_i \partial \gamma_j} \Big| \F_{t-1}\bigg).
\]
Under some regularity conditions (see Section \ref{s:asinf}),
\begin{equation}\label{ah}
 \frac{1}{n}H(\bs \gamma) - I^{(n)}(\bs \gamma) \overset{P}{\longrightarrow} 0 \quad \mbox{and} \quad    \frac{1}{n}K_n(\bs \gamma) - I^{(n)}(\bs \gamma) \overset{P}{\longrightarrow} 0, \quad \mbox{as} \quad n\to \infty.
\end{equation}
Furthermore, $I^{(n)}(\bs \gamma) \, {\longrightarrow} \,  I(\bs \gamma)$, where
\[
I(\bs \gamma) = \lim_{n\to \infty} I^{(n)}(\bs \gamma) = \lim_{n\to \infty} -\frac{1}{n}\E \bigg(\frac{\partial^2\ell(\bs \gamma)}{\partial \bs \gamma \partial \bs \gamma'} \bigg)
\]
which is the analogous of the $I_1(\bs \gamma)$ matrix for i.i.d. samples.

In order to derive $K_n$ for the model defined by \eqref{eq:dist_cond} and \eqref{e:gmu}, observe that the first derivative of the log-likelihood $\ell_t := \ell_t(\bs\gamma)$ with respect to $\gamma_j$ can be written as
\[
\frac{\partial \ell_t(\bs\gamma)}{\partial \gamma_j} = \frac{\partial \ell_t}{\partial \mu_t}
 \frac{\partial \mu_t}{\partial \gamma_j} +  \frac{\partial \ell_t}{\partial \varphi}\frac{\partial \varphi}{\partial\gamma_j} 
\]
so that
\begin{align*}
 \frac{\partial^2\ell_t(\bs\gamma)}{\partial \gamma_i \partial \gamma_j} 
 & =  \bigg[\frac{\partial^2 \ell_t}{\partial \mu_t^2}
 \frac{\partial \mu_t}{\partial \gamma_j} + \frac{\partial \ell_t}{\partial \mu_t}
  \frac{\partial }{\partial \mu_t}\bigg(\frac{\partial \mu_t}{\partial \gamma_j}\bigg) +  \frac{\partial^2 \ell_t}{\partial \mu_t\partial \varphi}\frac{\partial \varphi}{\partial\gamma_j} + \frac{\partial \ell_t}{\partial \varphi}\frac{\partial }{\partial \mu_t}\bigg(\frac{\partial \varphi}{\partial\gamma_j}\bigg)\bigg]\frac{\partial \mu_t}{\partial \gamma_i}\\
     & \quad \quad  \quad   + \bigg[\frac{\partial^2 \ell_t}{\partial\varphi\partial \mu_t}
 \frac{\partial \mu_t}{\partial \gamma_j} + \frac{\partial \ell_t}{\partial \mu_t}
  \frac{\partial }{\partial \varphi}\bigg(\frac{\partial \mu_t}{\partial \gamma_j}\bigg) +  \frac{\partial^2 \ell_t}{ \partial \varphi^2}\frac{\partial \varphi}{\partial\gamma_j} + \frac{\partial \ell_t}{\partial \varphi}\frac{\partial }{\partial \varphi}\bigg(\frac{\partial \varphi}{\partial\gamma_j}\bigg)\bigg]\frac{\partial \varphi}{\partial\gamma_i}.
\end{align*}
Since,
\[
 \frac{\partial \mu_t}{\partial \gamma_k}, \quad  \frac{\partial \varphi}{\partial\gamma_k}, \quad \frac{\partial }{\partial \mu_t}\bigg(\frac{\partial \mu_t}{\partial \gamma_k}\bigg), \quad \frac{\partial }{\partial \mu_t}\bigg(\frac{\partial \varphi}{\partial\gamma_k}\bigg),  \quad   \frac{\partial }{\partial \varphi}\bigg(\frac{\partial \mu_t}{\partial \gamma_k}\bigg), \quad  \mbox{and} \quad \frac{\partial }{\partial \varphi}\bigg(\frac{\partial \varphi}{\partial\gamma_k}\bigg)
\]
are all $\F_{t-1}$-measurable, it follows that
\[
\E\bigg(\frac{\partial \ell_t}{\partial \mu_t} \bigg| \F _{t-1}\bigg)= 0 \quad \mbox{and} \quad  \E\bigg(\frac{\partial \ell_t}{\partial \varphi} \bigg| \F _{t-1}\bigg)= 0.
\]
Hence
\begin{align*}
 [K_n]_{i,j}  & =   \sum_{t=1}^{n}\bigg\{\bigg[\E\bigg(\frac{\partial^2 \ell_t}{\partial \mu_t^2} \bigg| \F _{t-1} \bigg)
  \frac{\partial \mu_t}{\partial \eta_{t}} \frac{\partial \eta_{t}}{\partial \gamma_j} +   \E\bigg(\frac{\partial^2 \ell_t}{\partial \mu_t\partial \varphi} \bigg| \F _{t-1} \bigg)  \frac{\partial \varphi}{\partial \gamma_j}\bigg] \frac{\partial \mu_t}{\partial \eta_{t}} \frac{\partial \eta_{t}}{\partial \gamma_i}\\
     & \qquad \qquad     + \bigg[\E\bigg(\frac{\partial^2 \ell_t}{\partial\varphi\partial \mu_t} \bigg| \F _{t-1} \bigg)
  \frac{\partial \mu_t}{\partial \eta_{t}} \frac{\partial \eta_{t}}{\partial \gamma_j}+  \E\bigg(\frac{\partial^2 \ell_t}{ \partial \varphi^2} \bigg| \F _{t-1} \bigg) \frac{\partial \varphi}{\partial \gamma_j}\bigg] \frac{\partial \varphi}{\partial \gamma_i}\bigg\}
\end{align*}
and the conditional Fisher information matrix for $\bs\gamma$ is then given by
\begin{align}\label{eq:fisher}
K_n(\bs\gamma) := \left( \begin{array}{cccc}
K_{\bs\rho,\bs\rho}&K_{\bs\rho,\varphi}\\
K_{\bs\lambda,\bs\rho}& K_{\varphi,\varphi}
\end{array}\right),
\end{align}
with
\[
K_{\bs\rho,\bs\rho}= D'_{\bs \rho}T_1E_\mu T_1 D_{\bs \rho}, \quad
K_{\bs\rho,\varphi}=K_{\varphi,\bs\rho}'= D_{\bs \rho}' T_1E_{\mu\varphi}\bs 1_n   \quad \mbox{and} \quad
 K_{\varphi,\varphi}= \bs 1_n' E_\nu\bs 1_n
\]
where $D_{\bs\rho}$, $T_1$ and $\bs 1_n$ are the matrices and the vector defined in \eqref{eq:score} and  $E_\mu $, $E_{\mu\nu}$ and  $E_\nu$ are diagonal matrices for which the $(t,t)$th element is given by
\[
[E_\mu ]_{t,t} = -\E\bigg(\frac{\partial^2 \ell_t}{\partial \mu_t^2} \bigg| \F _{t-1} \bigg), \,\,
[E_{\mu\nu}]_{t,t} = -\E\bigg(\frac{\partial^2 \ell_t}{\partial\mu_t\partial \varphi} \bigg| \F _{t-1} \bigg) \,\, \mbox{and} \,\,
[E_\nu]_{t,t} = - \E\bigg(\frac{\partial^2 \ell_t}{ \partial \varphi^2} \bigg| \F _{t-1} \bigg).
\]

\section{Asymptotic theory and hypothesis testing}\label{s:asinf}

A rigorous asymptotic theory for the PMLE in the context of GARMA-like models the underlying distribution belongs to the canonical exponential family can be found in \cite{Fokianos1998,Fokianos2004}. Although the exponential family is broad enough to be useful in practice, this is still a limitation that must be observed in practice. For PTSR models, when the underlying distribution belongs to the canonical exponential family, the model falls into the context of \cite{Fokianos2004} and the asymptotic theory for the PMLE follows under assumptions A1 to A4 there stated.  Under those conditions, there exists a non-random information matrix, denoted by $I(\bs\gamma)$, such that
\begin{equation*}
\frac{K_n(\bs\gamma)}n\overset{P}{\underset{n\rightarrow\infty}\longrightarrow} I(\bs\gamma),
\end{equation*}
holds (in probability),  $I(\bs{\gamma})$ is positive definite and invertible matrix in an open neighborhood of the true parameter $\bs\gamma_0$. It can also be shown that the probability that a locally unique maximum partial likelihood estimator exists in a neighborhood of $\bs\gamma_0$ tends to one. Furthermore, the estimator is consistent
\[\widehat{\bs\gamma}_n\overset{P}{\underset{n\rightarrow\infty}\longrightarrow} \bs\gamma_0\]
asymptotically normal
\begin{equation}\label{an}
\sqrt{n}(\widehat{\bs\gamma}_n-\bs\gamma_0)\overset{d}{\underset{n\rightarrow\infty}\longrightarrow} N_{p+q+s+2}\big(\bs0,I(\bs\gamma_0)^{-1}\big),
\end{equation}
and \eqref{ah} holds.

For distributions that are not member of the canonical exponential family, a general asymptotic theory for the PMLE in the context of GARMA-like models is not available. We speculate that, under assumptions closely related to A1 to A4 in \cite{Fokianos2004}, the proofs presented in  \cite{Fokianos1998} can be adapted to provide asymptotic results similar to \eqref{an}, in a case by case fashion. However, we shall not pursuit this matter here.

\subsection{Confidence intervals and hypothesis testing inference}

Construction of asymptotic confidence intervals/regions and test statistics for hypothesis testing can be obtained using \eqref{an}. Let $\{Y_t\}_{t=1}^n$ be a sample from a PTSR model, $\gamma_{i}$ denote the $i$th component of the true parameter vector $\bs\gamma$ and let $\hat{\gamma}_i$ be its PMLE obtained from the sample. Let
$I(\hat{\bs\gamma})^{ij}$ denote the $(i,j)$th element of the inverse of the conditional information matrix \eqref{eq:fisher} evaluated at $\hat{\bs\gamma}\in\R^{p+q+s+2}$.

From \eqref{an}, we have
\begin{align*}
\frac{\hat{\gamma}_i - \gamma_i}{\sqrt{I(\hat{\bs\gamma})^{ii}}} \stackrel{d}{\longrightarrow} \mathcal{N}(0,1).
\end{align*}
Hence a $100(1-\alpha)\%$, $0 < \alpha < 1/2$, asymptotic confidence interval for $\gamma_i$ is given by
\begin{align*}
\left[\hat{\gamma}_i
- z_{1-\alpha/2} \sqrt{I(\hat{\bs\gamma})^{ii}};\hat{\gamma}_i
+ z_{1-\alpha/2} \sqrt{I(\hat{\bs\gamma})^{ii}}\right],
\end{align*}
where $z_{\delta}$ is the $\delta$-quantile of the standard normal distribution.

From \eqref{an} one can also derive asymptotic test statistics for hypothesis testing. Let $\gamma_{i}^0$ be a given hypothesized value for the true parameter $\gamma_{i}$ and consider the test
\[
\mathcal{H}_0:\gamma_{i}=\gamma_{i}^0 \quad \mbox{against} \quad \mathcal{H}_1:\gamma_{i} \neq \gamma_{i}^0.
\]
An asymptotic version for the signed square root of Wald's statistic can be obtained from \eqref{an} by considering
\begin{align*}
Z = \frac{\hat{\gamma}_i-\gamma_{i}^0}{\sqrt{I(\hat{\bs\gamma})^{ii}}}\,.
\end{align*}
Under $\mathcal{H}_0$, the distribution of $Z$ is approximately standard normal for large $n$. For details and a proof of this claim see \cite{Pawitan2001} and \cite{Fahrmeir1987}.

Versions for other well-known statistics such as the likelihood ratio, Rao's score, Wald's and the gradient statistics to perform more general hypothesis testing inference can also be derived from \eqref{an} in similar fashion. In large samples and under the null hypothesis, such statistics are approximately distributed as in the traditional i.i.d. case. More generally, for $k<p+q+s+2$, let $T:\R^{p+q+s+2}\rightarrow\R^k$  be a vector valued transformation such that its jacobian $\bs J(\bs\gamma)$ exists, is of full rank $k$ and it is a continuous function of $\bs\gamma$ in an open subset of $\Omega$. To test a composite hypothesis of the form
\[
\mathcal{H}_0:T(\bs\gamma)=\bs 0 \quad \mbox{versu} \quad \mathcal{H}_1:T(\bs\gamma)\neq \bs 0.
\]
we can use the traditional Wald's statistic, given by
\begin{equation*}
W=nT(\hat{\bs\gamma})'\big[\bs{J}(\hat{\bs\gamma})'I^{-1}(\hat{\bs\gamma})\bs{J}(\hat{\bs\gamma})\big]^{-1}T(\hat{\bs\gamma}).
\end{equation*}
Under $\mathcal{H}_0$ its distribution converges to a chi-square with the usual degrees of freedom.

\section{Diagnostic analysis and forecasting}\label{s:diag}

\subsection*{Model selection criteria}

Diagnostics in the context of PTSR models follow the usual procedures of GLM theory.
%

Model selection among several competing models may be based on the usual information criteria such as Akaike's (AIC), Schwartz's (SIC) and Hannan Quinn's (HQ) information criteria, respectively defined by
\[
{\rm AIC} = -2 \hat{\ell} +2s,\quad  {\rm SIC} = -2 \hat{\ell} +\log(n) s,\quad\mbox{and}\quad
{\rm HQ} = -2 \hat{\ell} + \log\left(\log(n)\right) k,
\]
where $k$ denotes the number of parameter in the fitted model and $\hat{\ell}$ is the log-likelihood function \eqref{eq:llk} evaluated at the PMLE. As usual, these criteria should be applied in combination with residual analysis, discussed next.

\subsection*{Residuals}

Residuals are extremely important in assessing the quality of fit from a model. There are several types of residuals that can be computed given a model. The most commonly applied ones are the simple residual given by $r_t=Y_t-\mu_t$ and the so-called quantile residuals defined by
\[e^{(q)}_t=\Phi^{-1}\big(F(Y_t|\F_{t-1})\big),\]
where $\Phi^{-1}$ denotes the standard normal quantile function. In the present framework, if the model is correctly specified, then $r_t$ should behave as a martingale difference (with respect to $\F_{t-1}$), while the quantile residuals should follow a standard normal distribution. These simple results are often applied in the construction of goodness-of-fit tests.

The literature related to testing the martingale difference hypothesis has grown significantly in the last decade and several tests and computational packages are available to perform such tests. For instance, \cite{kim2009} proposed the so-called wild bootstrap automatic variance ratio test. \cite{dominguez} proposed an approach based on the Cramer von Mises and Kolmogorov-Smirnov statistics to test the martingale difference hypothesis, which is called the Dom\'inguez-Lobato test. Finally, another approach based on the generalized spectral distribution function is presented in \cite{escanciano}. We refer the reader to the aforementioned papers for details. See also \cite{charles}, where the authors discuss finite sample performance of these methods. Such tests are readily available in most softwares. For instance in R \citep{R}, they are available in the package \texttt{vrtest} \citep{vrtest}.

In the present scenario, after we perform parameter estimation, we can obtain an estimate for the simple residuals. Then a martingale difference test can be applied to the estimated simple residuals, resulting in a goodness-of-fit test. As long as the second moment of the fitted model is finite, one can also apply a white noise test to the estimated simple residuals as, in this scenario, a martingale difference is, unconditionally,  a white noise.

When the model is correctly specified, the quantile residual should follow a standard normal distribution. Hence, testing the estimated quantile residual for normality can be used a goodness-of-fit test.
Another useful diagnostic tool is as follows. When $n$ is sufficiently large, the distribution of the residuals sample autocorrelation function at lag $h$,  $\hat{\rho}(h)$, is approximately normal with zero mean and constant variance $1/(n-m)$ \citep{Kedem2002, Anderson1942, Box2008}. The plots of the residual ACF with horizontal lines at $\pm 1.96/\sqrt{(n-m)}$ can be useful for assessing whether the residuals display white noise behavior~\citep{Kedem2002}. The traditional Ljung-Box test \citep{Ljung1978} based on the residual, to test the null hypothesis $\mathcal{H}_0:\rho(1)=\dots=\rho(l)=0$, for some $l>0$, the following test statistic can be used
\[Q= n(n+2) \sum_{i=1}^l \frac{\hat{\rho}(i)^2}{n-i}.\]
Under the null hypothesis and large $n$, $Q$ is approximately chi-squared distributed with $l$ degrees of freedom.

\subsection{Forecasting}

Upon applying the partial maximum likelihood estimators  in \eqref{e:gmu}, we can obtain the in-sample forecast, denoted by $\{\hat\mu_t\}_{t=1}^n$, and the $h$ steps ahead predicted values (out-of-sample forecast) for the conditional mean of a PTSR model, which we denote by  $\hat{\mu}_{n+h}=\hat{\mu}_{n}(h)$. We shall assume that the covariates $\bs{X}_t$, for $t=n+1,\dots,n+h$, are available or can be obtained.

Starting at $t=1$, we sequentially set
\[
\hat \eta_{t}  = \hat\alpha + \hat{\bs X}_t'\hat{\bs\beta} + \sum_{i=1}^p  \hat\phi_i\bigl[g_{1,2}(\hat Y_{t-i})-I_{X}\hat{\bs X}_{t-i}'\hat{\bs\beta}\bigr] + \sum_{k=1}^q \hat \theta_k \hat e_{t-k},
\]
with
  \[
    \hat Y_t = \begin{cases}
    0, & p = 0, \  t < 1,\\
    \frac{1}{p}\sum\limits_{i = 1}^p Y_i, & p > 0, \ t < 1,\\
   Y_t, & 1 \leq  t \leq n,\\
  \hat \mu_t, & t > n,
   \end{cases}\quad
   \hat{\bs X}_t' = \begin{cases}
   0, & pI_X = 0, \ t < 1,\\
\frac{1}{p}\sum\limits_{i = 1}^p \bs X_i', & pI_X >0, \ t < 1,\\
\bs X_t, & t \geq 1,
\end{cases}
\]
\[
\hat \mu_t  = g_1^{-1}(\hat \eta_t), \ t \geq 1,    \quad \mbox{and} \quad
  \hat e_t = \begin{cases}
   \hat Y_t - \hat \mu_t, & 1 \leq t \leq n,\\
  0, & \mbox{otherwise}.
   \end{cases}
  \]

\bibliographystyle{elsarticle-harv}
\bibliography{biblio}

\end{document}